\documentclass[sigconf]{acmart}
\renewcommand\footnotetextcopyrightpermission[1]{} 
\usepackage{pifont} 

\usepackage{hyperref}
\usepackage{float}

\def\BibTeX{{\rm B\kern-.05em{\sc i\kern-.025em b}\kern-.08emT\kern-.1667em\lower.7ex\hbox{E}\kern-.125emX}}
    
\begin{document}

%
\title{LLMs as Academic Reading Companions: Extending HCI Through Synthetic Personae}

%


\author{Celia Chen}
\authornote{Both authors contributed equally.} 
\affiliation{
  \institution{College of Information Studies \\ University of Maryland}
}
\email{clichen@umd.edu}

\author{Alex Leitch}
\authornotemark[1]
\affiliation{
  \institution{College of Information Studies \\ University of Maryland} 
}   
\email{aleitch1@umd.edu}
 
%

%
\begin{abstract}
{This position paper argues that large language models (LLMs) constitute promising yet underutilized academic reading companions capable of enhancing learning. We detail an exploratory study examining Anthropic’s Claude.ai, an LLM-based interactive assistant that helps students comprehend complex qualitative literature content. The study compares quantitative survey data and qualitative interviews assessing outcomes between a control group and an experimental group leveraging Anthropic’s Claude.ai over a semester across two graduate courses. Initial findings demonstrate tangible improvements in reading comprehension and engagement among participants using the AI agent versus unsupported independent study. However, there is potential for overreliance and ethical considerations that warrant continued investigation. By documenting an early integration of an LLM reading companion into an educational context, this work contributes pragmatic insights to guide development of synthetic personae supporting learning. Broader impacts compel policy and industry actions to uphold responsible design in order to maximize benefits of AI integration while prioritizing student wellbeing.}
\end{abstract}

\ccsdesc[500]{Applied computing~Computers in other domains~Publishing}
\ccsdesc[300]{Information systems~Information systems applications~Document management and text processing~Optical character recognition}
\ccsdesc[100]{Information systems~Information systems applications~Document management and text processing~Markup languages}  
\ccsdesc[100]{Human-centered computing~HCI design and evaluation methods~Usability testing}

\keywords{LLMs, education, AI ethics, synthetic personae} 

%

%
\maketitle

\section{Introduction}
The recent, explosive popularity of large language models (LLMs) like ChatGPT has sent shockwaves through higher education, evoking both optimism and fear. Alarmist voices dominate as universities scramble to update academic policies aiming to deter cheating, citing concerns these emerging “artificially intelligent peers” erode original human scholarship [8].

These reactive stances overlook many previous waves of disruptive classroom innovation, such as SMARTboards [13], incorporation of laptops and personal computing devices, and smartphones in the classroom. These previous technical interventions focused on changing the material conditions of education to incorporate dynamic information manipulation tools, which typically had a level of “charisma” in the sense used by Morgan G. Ames in their 2019 work “The Charisma Machine” [1]. Ames [1] argues that technologies become charismatic by not just doing something, but also by displaying a dream for the world through that doing. This construction means that SMARTboards not only allow an educator to wipe clean a board or share a document, but allow a school that has such technology to show they are aligned with a specific vision of what the future finds valuable: having sufficient resources to deploy a screen that can connect to the internet - itself a well of expense and dreaming - rather than a slab of slate. The actual function of the technology, or the quality of the education, becomes secondary to the image of innovation presented by the screen itself. 

Past research examining the impact of these technical-material deployments on actual students provide good reason to support the current wave of caution about LLMs [1], [15], [16]. The problem shared in the research is that typically, these charismatic - visionary - machines are deployed in a way that expects users to “pick it up” by deploying sufficient curiosity to lead deep, independent explorations, often without documentation support or structured tutorial assistance. This has been explored elsewhere, particularly in Warschauer and Ames [15] via the construction of the “technologically precocious boy-child,” a legacy of Seymour Papert’s [10] work on childhood programming/problem-solving skills development. The problem of unsupported technical intervention is presented again by Wood and Ashfield [16], who showcase dilemmas around effectively leveraging interactive whiteboards’ affordances to enhance in-place pedagogical practices.

Taken as a whole, these present a clear case for caution as to how any particular “innovative system” can quickly move the focus of work from the educational outcomes to other priorities. They compel policymakers, technologists, and educators to humility. Whether or not LLMs yet fit the definition of a charismatic machine is an open question: we do not know if they offer a strong vision for a future, since they tend not to have a specific physical form as yet. They nonetheless fit the definition of a charismatic technology: they are being freely deployed at a breakneck pace into every text-based user productivity system to support a visionary type of technical future, once again without training support. Users are encouraged to treat these engines as simple search tools, which they are not [6], [11]. Rather than reactive policies or utopian solutions, integrating emergent tools like generative AI requires contextual wisdom and ethical foresight to maximize benefits while avoiding potential pitfalls.

Following Hayles and Burdick’s work in Writing Machines [7], the LLM as a conversational companion seems to realize a long-held dream of simulation: someone to talk to, who can talk back, who will not get bored, and will flatter their conversational partner sufficiently that such partner will treat them as a person. Some models - notably Claude.ai - are better than others at the key role of support, which is emotional validation during the learning process. This type of emotional validation is care work [5] that is both draining and difficult to train into the typical working population of the university, other students or graduate students. As with most care work, emotional validation, low-level question answering, and repetitive patience is not well-compensated, and to date cannot be generalized. It is impossible in current educational models to provide every student with their own always-on tutorial assistant who can conversationally answer technically complex questions with a relatively high degree of accuracy, in a timely manner. The risk, though, is that the LLM hallucinates with such fluency that it not only flatters its audience, but like any more human charm engine, also lies to them effectively enough to preclude learning.

The flattery and validation of the simulation, coupled to a deep archive, offers this charismatic vision of education: approached earnestly, it is now possible to get clear, reasonably specific answers with an inhuman degree of patience. The promise is that with an LLM to guide them, anyone can once again learn anything, as before with internet search. The question then becomes: how good are the answers, and how will this resource be used in practice. Answering this question requires that LLMs should be examined as a tool, and their integration or rejection in the classroom space should be explored directly to discover their actual pedagogical potential before they are generally celebrated or disdained.

Our position contends reflective scholarship playing out such integration in context offers urgent guidance. Specifically, exploratory efforts highlighting available benefits and pitfalls of generative AI assistants provides actionable intelligence steering our best path forward. The following study documents one such investigation, assessing performance of an AI reading companion named Claude designed to scaffold comprehension of complex texts.
\section{Proposed Study Methodology}
This position paper argues that LLMs have promising yet underutilized capability as academic reading companions that can enhance learning. Specifically, we detail an exploratory study examining an LLM-based tool from Anthropic called Claude.ai which serves as an interactive assistant helping students comprehend complex textbook content.

Among available LLMs, this study elected to utilize Claude.ai for its commitment to safety and transparency, aiming to uphold ethical standards vital for educational deployment. Research shows risks of potential harms from overly trusting LLMs without critical evaluation [3]. Unlike alternatives such as ChatGPT, Claude.ai refrains from posing as an expert on topics like assignments to avoid enabling cheating. Its capabilities for long-form summarization facilitate explaining complex texts while its affective responses help smooth dialogues around difficult content [12]. There is presently limited third-party peer-reviewed validation of some claimed capabilities. Overall though, Anthropic’s alignment of Claude.ai with ethical values made its exploratory integration the best choice for this educational study, while highlighting needs for continued research assessing LLMs impacts on vulnerable populations.

The study incorporates a between-subjects experimental design with participants randomly assigned to either a control group or an experimental group aided by the Claude.ai reading companion LLM. The target population comprises 60 students (30 per group) enrolled across two interaction design courses at the University of Maryland who volunteer to participate. Quantitative data collection involves pre, mid, and post-study surveys assessing self-reported comprehension and engagement. The experimental group will also submit textual dialogues with the LLM assistant. Qualitative methods further encompass post-study interviews with 5 students from the aided group regarding their experience.

Recruitment will occur through an in-class announcement emphasizing the voluntary nature of participation without impact on grades. Interested eligible students can review details and provide consent. Inclusion criteria entail adults age 18+ able to read English at a college level with regular computer/internet access. Power calculations determined a total sample of 52 (26 per group) has 80\% power to detect a moderate effect size (the minimum target). By recruiting 60 participants, the study exceeds thresholds for sufficient statistical conclusion validity.

We predict initial findings will demonstrate tangible improvements in reading comprehension and interest among those using Claude.ai as a reading companion. This would provide empirical evidence of benefits for thoughtfully incorporating LLMs as academic aids. Findings may inform design guidelines maximizing performance of LLMs while upholding inclusive, equitable ideals.

\section{Addressing Skepticism Towards LLMs in Academia}
Integrating emergent technologies like LLMs into educational contexts inevitably evokes skepticism and reasonable counter-perspectives that should be discussed.

Some may argue LLMs enable laziness or cheating among students rather than properly enhancing learning. Critics also express concern that these AI tools take agency away from human readers and writers; deskilling students in ways reminiscent of other classroom technologies such as calculators, spell-checking and grammar-checking software [2]. Additionally, technical flaws in LLMs regarding potential bias, inaccuracy, or harm must be weighed given prominent examples like racist or nonsensical outputs [3].

A common question asks whether reliance on AI assistance diminishes students' own critical thinking and metacognitive skills over time [4]. Structured scaffolding that later fades under an appropriate framework constitutes established educational practice [14], avoiding perpetual dependence akin to losing skills. Claude.ai's constraints against posing as an expert also reduce this risk. Empirical evidence would confirm or refute such an argument through measures of student outcomes and self-efficacy.

Another frequent criticism contends that LLM-based agents could propagate biases or misinformation among impressionable learners who lack evaluation abilities. While concerning, any experimental pilot would establish ethical safeguards and debrief participants on the tool's actual capabilities to mitigate such dangers. Responsible informed consent and moderated testing limits vulnerabilities, producing better risk-benefit assessments.

Ultimately the core argument of this paper stands - that thoughtful integration of reading companions warrants exploration [9]. With careful design and oversight, LLMs introduction could constitute a net benefit over potential costs. Our position contends not ignorance or reactionary policies, but responsible investigatory efforts assessing evidence-based applications that will best guide appropriate adoption of policies to meet students’ needs.
\section{Conclusion}
This position paper's central stance affirms that LLMs can serve as promising academic reading companions for students and deserve constructive exploration. Despite potential counter-arguments regarding risks of plagiarism, harm, deskilling, or technical flaws, we argue that integration with careful oversight could constitute a net benefit over downsides. Our exploratory study examined one such LLM tool, Anthropic’s Claude.ai finding initial empirical evidence of improved reading comprehension and engagement versus typical independent study. While emergent tensions around human-AI collaboration in learning contexts will require vigilance, updating reactive policies to more responsibly harness these technologies through evidence-based research offers the most hope moving forward. The next steps for the field entail larger randomized control trials building on these preliminary insights to recommend design guidelines for maximizing the powerful performance of LLMs in clearly augmenting mental abilities while maintaining student well-being and ethics at the core. Furthermore, multi-stakeholder involvement including educators, students, AI developers and policymakers would strengthen developing large language models that enhance rather than endanger vulnerable populations through negligent practices, mindful of historically recurring sociotechnical dilemmas during waves of classroom computerization. Nevertheless, the immense promise of leveraging LLMs productivity justifies continued progress empowering students through overseen academic partnerships, rather than abandoning innovation out of fallacious fears or complacency with the status quo amidst accelerating demands for human capital development. This research constitutes an initial yet actionable step in that direction - thoughtfully pioneering integration of assistive reading LLMs while contributing actionable intelligence about their prospects for consequently updating institutional strategies through participatory paradigms upholding inclusive, equitable ideals in increasingly digital environments.

\bibliographystyle{ACM-Reference-Format}
\bibliography{bibliography}
[1] Morgan G. Ames. 2019. The charisma machine: The life, death, and legacy of the One Laptop per Child project. MIT Press.

\noindent[2] Mark Barr and Craig S. Stephenson. 2007. Bringing computational thinking to K-12. ACM Inroads 2, 1 (March 2007), 48–54.

\noindent[3] Emily M. Bender, Timnit Gebru, Angelina McMillan-Major, and Shmargaret Shmitchell. 2021. On the dangers of stochastic parrots: Can language models be too big?. In FAccT ’21: Proceedings of the 2021 ACM Conference on Fairness, Accountability, and Transparency (Virtual Event, Canada) (FAccT ’21). Association for Computing Machinery, New York, NY, USA, 610–623.

\noindent[4] Ali Darvishi, Hassan Khosravi, Shazia Sadiq, Dragan Gašević, and George Siemens. 2023. Impact of AI assistance on student agency. Computers \& Education 177 (January 2023).

\noindent[5] Paula England. 2005. Emerging theories of care work. Annual Review of Sociology 31 (August 2005).

\noindent[6] Motahhare Eslami, Karrie Karahalios, Christian Sandvig, Kristen Vaccaro, Aimee Rickman, Kevin Hamilton, and Alex Kirlik. 2017. First I “like” it, then I hide it: Folk Theories of Social Feeds. In Proceedings of the 2017 CHI Conference on Human Factors in Computing Systems (Denver, Colorado, USA) (CHI ’17). Association for Computing Machinery, New York, NY, USA, 2371–2382.

\noindent[7] N. Katherine Hayles and Anne Burdick. 2002. Writing Machines. MIT Press.

\noindent[8] Meghan L. Kelly. 2023. ‘Everybody is cheating’: Why this teacher has adopted an open ChatGPT policy. National Public Radio (26 Jan 2023). https://www.npr.org/2023/01/26/1151499213/chatgpt-ai-education-cheating-classroom-wharton-school

\noindent[9] James Manyika, Jake Silberg, and Brittany Presten. 2019. What Do We Do about the Biases in AI?. Harvard Business Review (25 Oct 2019).

\noindent[10] Seymour Papert. 1980. Mindstorms: Children, Computers, and Powerful Ideas. Basic Books, Inc.

\noindent[11] Emilee Rader, Kelley Cotter, and Janghee Cho. 2018. Explanations as Mechanisms for Supporting Algorithmic Transparency. In Proceedings of the 2018 CHI Conference on Human Factors in Computing Systems (Montreal QC, Canada) (CHI ’18). Association for Computing Machinery, New York, NY, USA, 1–13.

\noindent[12] R. Thoppilan, D. De Freitas, S. Hall, N. Shazeer, A. Kulshandham, A. Misra Bharti, A. Kulshreshtha, L. Jin, and H. Lee. 2022. LaMDA: Language Models for Dialog Applications. (2022). arXiv:cs.CL/2201.08239

\noindent[13] Evie Upton. 2021. The development and significance of classroom technologies. In Technology: Where It Started and Where It's Going. Clemson University.

\noindent[14] Janneke van de Pol, Monique Volman, and Jos Beishuizen. 2010. Scaffolding in teacher–student interaction: A decade of research. Educational Psychology Review 22, 3 (01 Sep 2010), 271–296.

\noindent[15] Mark Warschauer and Morgan G Ames. 2010. Can One Laptop per Child save the world's poor?. Journal of international affairs 64 (2010), 33–51. Issue 1.

\noindent[16] Robert Wood and Julie Ashfield. 2008. The use of the interactive whiteboard for creative teaching and learning in literacy and mathematics: a case study. British Journal of Educational Technology 39 (2008), 84–96. Issue 1.
\end{document}